\documentstyle[a4,12pt,axodraw]{article}
\catcode `@=11 \@addtoreset{equation}{section} \catcode `@=12
\begin{document}
\def\be{\begin{equation}}
\def\ee{\end{equation}}
\def\ba{\begin{array}{l}}
\def\ea{\end{array}}
\def\bea{\begin{eqnarray}}
\def\eea{\end{eqnarray}}
\def\eq#1{(\ref{#1})}
\def\fig#1{Fig \ref{#1}} 
\def\wgnc{\bar{\wedge}}
\def\del{\partial}
\def\der{\overline \del}
\def\wg{\wedge}
\def\bull{$\bullet$}
\def\gap{\vspace{10ex}}
\def\tgap{\vspace{3ex}}
\def\sgap{\vspace{5ex}}
\def\lgap{\vspace{20ex}}
\def\half{\frac{1}{2}}
\def\pto{\vfill\eject}
\def\gst{g_{\rm st}}
\def\tC{{\widetilde C}}
\def\z{{\bar z}}
\def\o{{\cal O}}
\def\J{{\cal J}}
\def\S{{\cal S}}
\def\X{{\cal X}}
\def\N{{\cal N}}
\def\A{{\cal A}}
\def\H{{\cal H}}
\def\D{{\tilde D}}
\def\d{{\cal D}}
\def\re#1{{\bf #1}}
\def\nn{\nonumber}
\def\nl{\hfill\break}
\def\ni{\noindent}
\def\bibi{\bibitem}
\def\c#1{{\hat{#1}}}
\def\eps{{\epsilon}}
\pretolerance=1000000
\begin{flushright}
KEK/TH/773\\
June 2001\\
\end{flushright}
\begin{center}
\vspace{2 ex}
{\large\bf Non-commutative Gauge Theory, \\
Open Wilson Lines and Closed Strings} 
\\
\vspace{3 ex}
Avinash Dhar $^{1*}$ and Yoshihisa Kitazawa $^{\dagger}$ \\
~\\
{\sl Laboratory for Particle and Nuclear Physics,}\\
{\sl High Energy Accelerator Research Organization (KEK),}\\
{\sl Tsukuba, Ibaraki 305-0801, JAPAN.} \\

\vspace{10 ex}
\pretolerance=1000000
\bf ABSTRACT\\
\end{center}
\vspace{1 ex} 

A recently proposed connection between closed string field and an open
Wilson line defined on an arbitrary contour is further explored here. We
suggest that reparametrization invariance of a Wilson line is the
principle which determines the coupling of non-commutative gauge
theory/matrix model to the modes of the closed string. An analogue of
the level matching condition on the gauge theory/matrix model
operators emerges quite naturally from the cyclic symmetry of the
straight Wilson line.  We show that the generating functional of
correlation functions of these operators has the space-time gauge
symmetry that one expects to find in closed string field theory.  We
also identify an infinite number of conserved operators in gauge
theory/matrix model, the first of which is known to be the conserved
stress tensor.

\vfill
\hrule
\vspace{0.5 ex}
\leftline{$^1$ On leave from Dept of Theoretical Phys, Tata Institute, Mumbai
400005, INDIA.}  
\leftline{$^*$ adhar@post.kek.jp}
\leftline{$^\dagger$ kitazawa@post.kek.jp}

\clearpage

\vspace{8ex}

\section{Introduction} 

The subject of the coupling of non-commutative branes to closed
strings in the bulk has been of much recent activity
\cite{DR,GHI,L,DT,OO,LM} \footnote{Other aspects of open Wilson lines
  were studied in \cite{AD,RU,DW,RR,DK2,BL,O}.}. At low energies, the
fluctuations of non-commutative branes are controlled by a
non-commutative gauge theory. So at low energies, the question of
coupling of non-commutative branes to closed strings is equivalent to
identifying gauge-invariant operators in the non-commutative gauge
theory which couple to the different modes of the closed string.
Gauge-invariant operators which couple to the tachyon and the
massless modes of the closed string are now known and these consist of
local operators smeared along straight Wilson lines \cite{IIKK1}.
Higher bulk modes are expected to follow a similar pattern.

It was pointed out in \cite{DK1} that the non-commutative gauge theory
operators dual to the tachyon and the massless modes in the case of
the bosonic string could be obtained as the first two terms in a
harmonic expansion of a Wilson line, based on a generic contour,
around a certain straight line contour. As was suggested in
\cite{DK1}, it seems natural to guess that the higher terms in the
expansion give rise to gauge theory operators that couple to higher
modes of the closed bosonic string. One may thus think of a Wilson
line based on an arbitrary contour as the gauge theory operator dual
to a bulk closed string. The purpose of the present work is to give
concrete evidence for this proposal. Specifically what we show is that
the operators that appear in the harmonic expansion of a Wilson line
satisfy a kind of level matching condition and are in one-to-one
correspondence with the modes of the closed bosonic string. Moreover,
the corresponding source terms in the action have the linearized gauge
symmetries expected in closed bosonic string field theory. Explicit
disc amplitude calculations in an appropriate zero slope limit
\footnote{In this limit, described in section 5, one gets leading
  terms in the gauge-invariant operators that couple to the
  corresponding modes of the closed string. It is these leading terms
  that are reproduced in the harmonic expansion of a Wilson line based
  on an arbitrary curve.} confirm the general structure of the
operators.

The plan of this paper is as follows. In the next section we discuss
the above mentioned harmonic expansion of an arbitrary Wilson line.
We suggest that the gauge theory operators that appear in the
expansion couple to the different modes of closed string. We show that
the cyclic symmetry of a straight Wilson line ensures a condition on
these operators which is similar to the level matching condition for
closed string modes.  Thus, the gauge-theory operators we get in this
way are in one-to-one correpondence with the modes of a closed string.
In section 3 we discuss reparametrizations of the arbitrary contour
underlying the Wilson line.  This leads to reparametrizations of the
straight line contour on which the operators involved in the expansion
of the Wilson line are based, and hence these operators change under
reparametrizations. However, we show that the correlation functions of
these operators remain unchanged because of momentum conservation. In
section 4 we discuss these operators from the point of view of the
matrix model underlying non-commutative gauge theory. The structure of
these operators is much simpler in the matrix model setting and we are
able to discuss some general features. In particular, we identify an
infinite number of conserved operators, the first of which is known to
be stress tensor of the matrix model. In section 5 we calculate some
disc amplitudes to provide further evidence that the operators that
appear in the harmonic expansion of an arbitrary Wilson line couple to
the modes of the closed string. In section 6 we discuss the symmetries
of the generating functional of the correlation functions of these
operators.  We show that reparametrization invariance of the original
Wilson line ensures that this generating functional has the space-time
gauge symmetries which one expects to find in closed string field
theory. We conclude in section 7 with a discussion of our results and
some open questions.

\section{Open Wilson lines and closed string modes}

In ordinary gauge theories, a generic gauge-invariant observable is
provided by an arbitrary closed Wilson loop. Non-commutative gauge
theories have more general gauge-invariant observables, defined on
open contours. Roughly speaking, these gauge-invariant observables can
be written as Fourier transforms of open Wilson lines. In the operator
formalism a generic open Wilson line is given by the following
expression
\bea 
W_C[y]={\rm Tr}\bigg({\rm P~exp}\{i\int_C d\sigma \ \del_\sigma y_\mu(\sigma)
A_\mu(\hat x+y(\sigma))\} \ e^{ik.\hat x} \bigg),
\label{twoone}
\eea
where
\bea
[{\hat x}_\mu, {\hat x}_\nu]=i\theta_{\mu\nu},
\label{twotwo}
\eea
and the trace in (\ref{twoone}) is over both the gauge group and the
operator Hilbert space \footnote{We will use the operator formulation
  throughout this paper. We are working with the fully non-commutative
  Euclidean case. See \cite{DK1} for details of our notation and
  conventions.}. The open Wilson line given by the above expression is
gauge-invariant provided the momentum $k^\mu$ associated with it is
fixed in terms of the straight line joining the end points of the path
$C$, given by $y^\mu(\sigma)$ where $0 \leq \sigma \leq 1$, by the
relation
\bea
y_\mu(1)-y_\mu(0)=\theta_{\mu\nu}k_\nu.
\label{twothree}
\eea
The contour $C$ is otherwise completely arbitrary. This condition may be
regarded as a boundary condition on the contours involved. A generic
contour with this boundary condition may be parametrized as
$y(\sigma)=\sigma(\theta k)+y'(\sigma)$, where $0 \leq \sigma \leq 1$
and $y'(\sigma)$ satisfies {\it periodic} boundary conditions. Thus
the freedom contained in a generic Wilson line is exactly what is
needed to describe a closed string!

In fact, this line of reasoning can be taken further. Let us confine
our attention to smooth contours, with the additional condition that the
tangents to the contour at the two ends are equal. In this case we may
parametrize the contours as
\bea
y(\sigma) &=& y_0(\sigma)+\theta w(\sigma), \nonumber \\
y_0(\sigma) &=& y_0(0)+\sigma l, \ l \equiv \theta k \nonumber \\
w(\sigma) &=& \sum_{n=1}^\infty (\alpha_n \ e^{-2\pi i n \sigma}+
\alpha_n^* \ e^{2\pi i n \sigma}).
\label{twofour}
\eea
In the above parametrization the factor of $\theta$ in front of the
periodic function $w_{\mu}(\sigma)$ has been chosen for later
convenience. Also, we choose the constant $y_0(0)$ such that $w(\sigma)$
contains no zero mode.

Let us now expand the Wilson line, based on the given contour, around
the Wilson line based on the straight line contour $C_0$ given by
$y_0(\sigma)$. This gives
\bea 
W_C[y] &=& W_{C_0}[y_0]+\int_0^1 d\sigma_1 \ (\theta w(\sigma_1))_\mu
\bigg({\delta  W_C[y] \over \delta y_\mu(\sigma_1)} \bigg)_{y=y_0} \nonumber \\
&& +{1 \over 2!}\int_0^1 d\sigma_1 \int_0^1 d\sigma_2 \ 
(\theta w(\sigma_1))_\mu \ 
(\theta w(\sigma_2))_\nu \bigg({\delta^2 W_C[y] \over \delta y_\mu(\sigma_1)
\delta y_\nu(\sigma_2)} \bigg)_{y=y_0}+\cdots
\label{twofive}
\eea
The first term in the above equation is known to be the
non-commutative gauge theory operator dual to the bulk closed string
tachyon. The second term vanishes, since $\bigg({\delta W_C[y] \over
\delta y_\mu(\sigma)} \bigg)_{y=y_0}$ is independent of $\sigma$,
which can be easily verified using the cyclic symmetry of a straight
Wilson line, and since $w(\sigma)$ has no zero mode. The first
non-trivial contribution comes from the third term. Using the identity
\bea
{\delta^2 W_C[y] \over \delta y_\mu(\sigma_1) \delta y_\nu(\sigma_2)} 
&=& {\rm Tr} \bigg[\bigg(\hat {{\cal U}_C}(0, \sigma_1)(i\del_{\sigma_1} 
y_\lambda(\sigma_1) \hat F_{\mu\lambda}(\hat x+y(\sigma_1)))
\hat {{\cal U}_C}(\sigma_1, \sigma_2) \nonumber \\
&& \hspace{5 ex} \times (i\del_{\sigma_2} y_\rho(\sigma_2) 
\hat F_{\nu\rho}(\hat x+y(\sigma_2)))\hat {{\cal U}_C}(\sigma_2, 1) 
\ e^{ik.\hat x} \theta(\sigma_2-\sigma_1) \nonumber \\
&& \hspace{5 ex} +(\sigma_1 \leftrightarrow \sigma_2, \mu 
\leftrightarrow \nu) \bigg) \nonumber \\ 
&& \hspace{5 ex} + \hat {{\cal U}_C}(0, \sigma_1)
(i\del_{\sigma_1} y_\lambda(\sigma_1)\delta(\sigma_1-\sigma_2) 
\hat D_\nu \hat F_{\mu\lambda}(\hat x+y(\sigma_1)))
\hat {{\cal U}_C}(\sigma_1, 1) \ e^{ik.\hat x} \nonumber \\
&& \hspace{5 ex} + \hat {{\cal U}_C}(0, \sigma_1)
(i \del_{\sigma_1}\delta(\sigma_1-\sigma_2)\hat F_{\mu\nu}(\hat x+y(\sigma_1)))
\hat {{\cal U}_C}(\sigma_1, 1) \ e^{ik.\hat x} \bigg]
\label{twosix}
\eea 
we may rewrite this term as
\bea
&& \theta_{\mu\mu'} \ \theta_{\nu\nu'} \bigg[
\int_0^1 d\sigma_1 \int_0^1 d\sigma_2 \ w_{\mu}(\sigma_1) \ 
w_{\nu}(\sigma_2) \ 
{\rm Tr} \bigg(\hat {\cal U}_{C_0}(0, \sigma_1) \ (il_\lambda 
\hat F_{\mu'\lambda}(\hat x+y_0(\sigma_1))) \ 
\hat {\cal U}_{C_0}(\sigma_1, \sigma_2) \nonumber \\
&& \hspace{15 ex} \times (il_\rho \hat F_{\nu'\rho}(\hat x+y_0(\sigma_2))) \  
\hat {\cal U}_{C_0}(\sigma_2, 1) \ e^{ik.\hat x} \ \theta(\sigma_2-\sigma_1)+
(\sigma_1 \leftrightarrow \sigma_2, \mu' \leftrightarrow \nu') \bigg) 
\nonumber \\
&& \hspace{15 ex} + \int_0^1 d\sigma_1  \ w_{\mu}(\sigma_1) \ 
w_{\nu}(\sigma_1) \ 
{\rm Tr} \bigg(\hat {\cal U}_{C_0}(0, \sigma_1) \ (il_\lambda 
\hat D_{\nu'} \hat F_{\mu'\lambda}
(\hat x+y_0(\sigma_1))) \ \hat {\cal U}_{C_0}(\sigma_1, 1) \ e^{ik.\hat x}
\bigg) \nonumber \\
&& \hspace{15 ex} + \int_0^1 d\sigma_1 \ w_{\mu}(\sigma_1) \ i\del_{\sigma_1}
w_{\nu}(\sigma_1) \ {\rm Tr} \bigg(\hat {\cal U}_{C_0}(0, \sigma_1) \ 
\hat F_{\mu'\nu'}(\hat x+y_0(\sigma_1)) \ 
\hat {\cal U}_{C_0}(\sigma_1, 1) \ e^{ik.\hat x} \bigg) \bigg], \nonumber \\
\label{twoseven}
\eea
where $\hat {\cal U}_C(\sigma_1, \sigma_2)$ is the path-ordered phase
factor, running along the contour $C$, from the point $\sigma_1$ to
$\sigma_2$ \footnote{In this notation $W_C[y]= {\rm Tr}(\hat {\cal
  U}_C(0, 1) \ e^{ik.\hat x})$.}.

One crucial ingredient that we will now use is the cyclic symmetry of
the Wilson line with the contour a straight line. It can be seen from
this symmetry that the operator appearing inside the parametric
integration in the first term in (\ref{twoseven}) depends only on the
difference $(\sigma_1-\sigma_2)$ while the operators appearing in the
other two terms are independent of $\sigma_1$. This ensures that when we 
substitute for $w(\sigma)$ from (\ref{twofour}) in it we get an
expression of the form 
\bea 
\sum_{n=1}^\infty \alpha_{n\mu} \alpha_{n\nu}^* 
O^{(n)}_{\mu\nu}(k)
\label{twoeight}
\eea
which involves the modes $\alpha$ and $\alpha^*$ at the {\it
same} mode number. Here $O^{(n)}_{\mu\nu}(k)$ are gauge theory
operators which can be easily worked out from the expressions given
above. For example, for $n=1$ the operator symmetric in the indices
$\mu, \nu$ turns out to be precisely the operator that has been
identified in \cite{OO} as dual to the bulk graviton in the bosonic
string. The last term in (\ref{twoseven}) is purely antisymmetric in
the indices $\mu, \nu$ and hence at the $n=1$ level it contributes
only to the operator dual to the bulk antisymmetric tensor field.

The precise form of the gauge theory operators dual to the higher
modes of the bosonic string has not been worked out, but the general
form is expected to be similar to that for the tachyon and the
massless modes. It is tempting to propose that these also appear in
the harmonic expansion (\ref{twofive}). Analogous to (\ref{twoeight}),
a generic term in the expansion (\ref{twofive}) has the form
\bea
\sum_{n_1,n_2, \cdots=1}^\infty \ \sum_{ m_1,m_2, \cdots=1}^\infty
\alpha_{n_1\mu_1} \alpha_{n_2\mu_2} \cdots \
\alpha_{m_1\nu_1}^* \alpha_{m_2\nu_2}^* \cdots \
O^{n_1,n_2,... \ ; \ m_1,m_2,...}_{\mu_1,\mu_2,... \ ; \ \nu_1,\nu_2,...}(k).
\label{twonine}
\eea
The sum over the mode numbers in the above expression is not free but
is constrained to satisfy $\sum_j n_j=\sum_j m_j$. As in the above
example of $n=1$, this condition follows from the cylic symmetry of
the straight Wilson line and is very similar to the level matching
condition that the physical states of closed string satisfy,
$\alpha_n$ and $\alpha_n^*$ being analogues of the two chiral modes of
the closed string. Thus the gauge theory operators that appear in the
harmonic expansion (\ref{twofive}) are in one-to-one correspondence
with the modes of closed bosonic string. Our proposal has, therefore,
passed a crucial consistency check.

\section{Straight Wilson lines and reparametrization invariance}

A Wilson line operator based on any contour, by its definition, must
not depend on any particular parametrization used for the contour.
Since this is also true of the Wilson line based on the generic
contour, $W_C[y]$, this means that the sum on the right hand side of
(\ref{twofive}) should be independent of the parametrization used for
the contour. The individual terms in the sum, however, are {\it not}
required to satisfy this condition and may change under
reparametrizations. Thus, in general the gauge theory operators
appearing in (\ref{twofive}) transform under reparametrizations. This
would have been a problem for our proposal, but it turns out that the
{\it correlation functions} of the operators remain unchanged, even
though the operators themselves change. Let us see this in some detail.

Let us consider a different parametrization of the contour $C$ given
by $y'(\sigma'), 0 \leq \sigma' \leq 1$, with the boundary condition
as before $y'(1)-y'(0)=l$. The harmonic expansion is now to be done
around the corresponding straight line contour. The analogue of
(\ref{twofour}) is \bea
y'(\sigma') &=& y_0'(\sigma')+\theta w'(\sigma'), \nonumber \\
y_0'(\sigma') &=& y_0'(0)+\sigma' l, \nonumber \\
w'(\sigma') &=& \sum_{n=1}^\infty (\alpha_n' \ e^{-2\pi i n \sigma'}+
\alpha_n'^* \ e^{2\pi i n \sigma'}),
\label{threeone}
\eea
where $\alpha_n'$ are the modes of $w'(\sigma')$ which, as before, has
no zero mode. For an infinitesimal reparametrization
$\sigma'=\sigma+\epsilon(\sigma)$, to the lowest order in  $\epsilon$, 
using $y'(\sigma')=y(\sigma)$ we get
\bea
y_0'(0)=y_0(0)-\epsilon_0 l-2 \pi i \theta \sum_{n=1}^\infty n(\epsilon_n
\alpha_n^*-\epsilon_n^* \alpha_n) 
\label{threetwo}
\eea
and
\bea
w'(\sigma)=w(\sigma)-\epsilon(\sigma)(k+\del_\sigma w(\sigma)).
\label{threethree}
\eea
The last equation holds only for non-zero modes with respect to $\sigma$, 
the zero mode of the product appearing in the second term on the right 
hand side being excluded from its consideration. Also, we have used the 
mode expansion
$\epsilon(\sigma)=\epsilon_0+ \sum_{n=1}^\infty (\epsilon_n e^{-2 \pi
i n \sigma}+ \epsilon_n^* e^{2 \pi i n \sigma})$ \footnote{Note that
for $\sigma'$ to range from $0$ to $1$ as $\sigma$ changes over the same 
range, we need $\epsilon(0)=\epsilon(1)=0$. This condition may be used
to express the zero mode $\epsilon_0$ in terms of the other modes of
$\epsilon(\sigma)$. Thus, $\epsilon_0$ is not an independent parameter.}.

The harmonic expansion of the original Wilson line around the new
straight line contour, $y_0'(\sigma')$, is similar to (\ref{twofive}),
\bea 
W_C[y'] &=& W_{C_0'}[y_0']+\int_0^1 d\sigma_1' \ (\theta w'(\sigma_1'))_\mu
\bigg({\delta  W_C[y'] \over \delta y'_\mu(\sigma_1')} \bigg)_{y'=y_0'} 
\nonumber \\
&+& {1 \over 2!}\int_0^1 d\sigma_1' \int_0^1 d\sigma_2' \ 
(\theta w'(\sigma_1'))_\mu \ 
(\theta w'(\sigma_2'))_\nu \bigg({\delta^2 W_C[y'] \over 
\delta y'_\mu(\sigma_1')
\delta y'_\nu(\sigma_2')} \bigg)_{y'=y_0'}+\cdots.
\label{threefour}
\eea
Since $\sigma_1'$, $\sigma_2'$, etc. are integration variables, we can
drop the primes on them. Because of this the gauge theory operators
involved in the above expansion differ from those involved in
(\ref{twofive}) only in that the starting point of the present
straight line is shifted with respect to the previous one by the
amount given in (\ref{threetwo}). This difference can be removed by
the shift $\hat x \rightarrow (\hat x + \epsilon_0 l+2 \pi i
\sum_{n=1}^\infty n(\epsilon_n \alpha_n^*-\epsilon_n^* \alpha_n))$ at
the expense of gaining the universal momentum-dependent phase 
\bea
e^{i l.\sum_{n=1}^\infty 2 \pi in(\alpha_n \epsilon_n^*-
\alpha_n^* \epsilon_n)}.
\label{threefoura}
\eea 
Thus, in terms of the modes $\alpha_n'$ the above expansion looks like
\bea
\sum_{n_1,n_2, \cdots=1}^\infty \ \sum_{ m_1,m_2, \cdots=1}^\infty
\alpha_{n_1\mu_1}' \alpha_{n_2\mu_2}' \cdots \
\alpha_{m_1\nu_1}'^* \alpha_{m_2\nu_2}'^* \cdots \
O'^{n_1,n_2,... \ ; \ m_1,m_2,...}_{\mu_1,\mu_2,... \ ; \ \nu_1,\nu_2,...}(k)
\label{threefive}
\eea
where the new operators $O'$ differ from the operators $O$ appearing
in (\ref{twonine}) only by the above phase.

One consequence of the above is that the correlation functions of the
individual gauge theory operators, which enter the harmonic expansion
of the Wilson line, are identical in any parametrization of the
contour $C$.  This is ensured by momentum conservation. It provides
another consistency check for our proposal to identify these operators
as dual to closed string modes.  Since reparametrizations act in such
a seemingly trivial fashion on the individual operators $O$, but since
the modes $\alpha_n'$ are non-trivially related to $\alpha_n$ by
(\ref{threethree}), one might ask how the sum in (\ref{threefive})
manages to be equal to (\ref{twonine}), as it must.  To see how this
happens it is instructive to study an example in detail, which is what
we will do now.

Let us consider the term in (\ref{threefive}) which correponds to the
antisymmetric tensor field. The piece in this operator which is linear 
in field strength is
\bea 
{i \over 2!} 2 \pi i \ \theta_{\mu\lambda} \theta_{\nu\rho} 
(\alpha_{1\mu}' \alpha_{1 \nu}'^*-\alpha_{1 \mu}'^* \alpha_{1 \nu}') \ 
{\rm Tr} \bigg( F_{\lambda\rho}(\hat x+y_0'(0)) \ \hat {\cal U}_{C_0'}(0,
1) \ e^{ik.\hat x} \bigg)
\label{threesix}
\eea
Using 
\bea
\alpha_1'=\alpha_1-\epsilon_1 k+2 \pi i\epsilon_0\alpha_1+
2 \pi i \sum_{n=1}^\infty 
((n+1)\epsilon_n^*\alpha_{n+1}-n\epsilon_{n+1}\alpha_n^*),
\label{threeseven}
\eea
which can be obtained from (\ref{threethree}), in (\ref{threesix}), we
find that it has the following extra terms over the corresponding term in 
(\ref{twonine})
\bea
&& 2 \pi i \theta_{\rho\mu}(\alpha_{1\mu}\epsilon_1^*-
\alpha_{1\mu}^*\epsilon_1)
\ {\rm Tr} \bigg(il_\lambda F_{\lambda\rho}(\hat x+y_0'(0))
\ \hat {\cal U}_{C_0'}(0, 1) \ e^{ik.\hat x} \bigg) \nonumber \\
&& \hspace{10 ex} + {\rm terms} \ {\rm quadratic} \ {\rm in} \ 
{\rm the} \ {\rm modes}.
\label{threeeight}
\eea
Now, it is easy to see that 
\bea
{\rm Tr} \bigg(il_\lambda F_{\lambda\rho}(\hat x+y_0'(0))
\ \hat {\cal U}_{C_0'}(0, 1) \ e^{ik.\hat x} \bigg)
&=& -{\delta  W_{C_0'}[y_0'] \over \delta y_{0\rho}'(0)} \nonumber \\
&=& ik_\rho \ W_{C_0'}[y_0'],
\label{threenine}
\eea
where in the first line above the variation with respect to $y_0'(0)$
is done keeping $k$ fixed. Using this in (\ref{threeeight}) and
retaining terms upto only first order in $\epsilon$, we get
\bea
&& -il.\{2 \pi i (\alpha_1\epsilon_1^*-\alpha_1^*\epsilon_1)\}
\ W_{C_0}[y_0]+ {\rm terms} \ {\rm quadratic} \ {\rm in} \ 
{\rm the} \ {\rm modes}.
\label{threeten}
\eea 
Now, the tachyon term in (\ref{threefive}) is $W_{C_0'}[y_0']$ which,
as discussed above, differs from $W_{C_0}[y_0]$ by the phase in
(\ref{threefoura}). To lowest order in $\epsilon_1$ this gives an
extra term which precisely cancels the extra contribution linear in
modes we have found above in (\ref{threeten}). This example
illustrates how the extra phase and the change in the modes due to a
reparametrization compensate each other. We expect this to happen for
all the other terms as well since this is required by the
reparametrization invariance of the Wilson line $W_C$. A general proof 
will be given in the next section where we discuss the connection with
the underlying matrix model. 

\section{Relation with matrix model}

It is useful to relate the above considerations to an underlying
bosonic matrix model \footnote{The relevant matrix model is the IIB
  type \cite{IKKT}. Possible existence of bosonic M theory has been
  speculated in \cite{B,FFZ,R,HS}.}.  The Wilson line of the
non-commutative gauge theory discussed above may be related to the
Wilson line of this matrix model,

\bea
{\tilde W}_C[y]={\rm Tr}\bigg({\rm P~exp}\{-i\int d\sigma \ \del_\sigma 
y_\mu(\sigma) \theta^{-1}_{\mu\nu} X_\nu\}\bigg),
\label{twoten}
\eea
using the by now familiar procedure of expanding around the appropriate 
brane solution, $X_\mu=\hat x_\mu -\theta_{\mu\nu} A_\nu(\hat x)$, 
where $\hat x_\mu$ satisfies the commutation relation given in (\ref{twotwo}).
It turns out that ${\tilde W}_C$ is not exactly the same as $W_C$ of 
(\ref{twoone}). However, the difference is only a phase which depends 
on the contour. That is \footnote{Similar expressions have appeared earlier
in \cite{O,I}. Such an expression is naturally obtained in the boundary
state formalism.},
\bea
{\tilde W}_C[y]= e^{i\phi_C[y]} \ W_C[y],
\label{twoeleven}
\eea
where
\bea
\phi_C[y]={1 \over 2}k.y(0)+{1 \over 2}\int^1_0 d\sigma \ y(\sigma)\theta^{-1}
\del_\sigma y(\sigma).
\label{twotwelve}
\eea  

An interesting aspect of above the phase can be seen by using the
parametrization (\ref{twofour}) in it. This gives
\bea
\phi_C[y]=k.(y(0)-\theta w(0))-{1 \over 2} \int^1_0 d\sigma \ w(\sigma)\theta 
\del_\sigma w(\sigma).
\label{twothirteen}
\eea
Using $y(0)=y_0(0)+\theta w(0)$ in this, substituiting in
(\ref{twoeleven}) and absorbing part of the the phase,
$e^{ik.y_0(0)}$, in the Wilson line by shifting to origin the starting
point of the corresponding straight line contour, we get
\bea
{\tilde W}_C[y]= e^{-{i \over 2}\int^1_0 d\sigma \ 
w(\sigma)\theta \del_\sigma w(\sigma)} \ W_k[w].
\label{twofourteen}
\eea
Here, we have used the notation $W_k[w]$ to indicate that the straight
line path involved in the Wilson line is now given simply by $\sigma
l$, the dependence on the starting point $y_0(0)$ having disappeared
from it. 

The operator $W_k[w]$ is {\it not} reparametrization invariant,
though, of course, the entire right hand side of (\ref{twofourteen})
does have this property.  This is because the original Wilson line
operator $W_C[y]$ and the phase $\phi_C[y]$ are independently
reparametrization invariant \footnote{Note that the phase
  $e^{-ik.y(0)}$ is reparametrization invariant.}. Therefore, a
physically meaningful harmonic expansion of the Wilson line that
naturally appears in the matrix model involves expanding $W_k[w]$
together with the $w$-dependent phase in (\ref{twofourteen}). This
gives gauge theory operators that are different from the operators
that appear in the expansion (\ref{twonine}).  One of the differences
is that the $w$-dependent phase in (\ref{twofourteen}) contributes
additional terms to the various operators. For example, at $n=1$ the
linear field strength piece in the operator that couples to the
antisymmetric tensor is changed from $F_{\mu\nu}(\hat x+y_0(0))$ to
$(F_{\mu\nu}(\hat x)+{\theta^{-1}}_{\mu\nu})$, attached to the
straight Wilson line. The other difference, which we have already
mentioned above, is that the straight line contour on which the new
operators are based is determined solely by the momentum $k$.  This
has the important consequence that the new operators are invariant
under reparametrizations, since the only source of reparametrization
dependence in the operators that appear in (\ref{twonine}) is the
starting point of the straight line contour, $y_0(0)$.  Since the
modes of the periodic function $w$ change under reparametrizations,
one might wonder as to how the sum in (\ref{twonine}) manages to be
reparametrization invariant, which it must be since the Wilson line is
by definition reparametrization invariant. Later on in this section we
will see how this happens.

Instead of doing a harmonic expansion of the right hand side of
(\ref{twofourteen}), one could directly expand (\ref{twoten}) itself.
This gives matrix model expressions for the gauge theory operators
involved in the harmonic expansion. These expressions are much simpler
than the expressions that appear in, for example (\ref{twoseven}),
and the general structure of the operators is more transparent. So it
is useful to discuss the harmonic expansion of (\ref{twoten}), which
is what we shall do now.

One property of the matrix model expression for the Wilson line is
that it depends only on the parametric derivative of the function
$y(\sigma)$ that characterizes the contour. Because of this, its
harmonic expansion is most easily obtained in terms of the derivative
of the periodic function $w(\sigma)$. In fact, it is easy to see that
\bea
{\tilde W}_C[y] &=& {\rm Tr}(e^{ik.X})+\int d\sigma \ 
i\del_\sigma w_\mu(\sigma) \ {\rm Tr}(e^{i\sigma k.X} \ X_\mu \ 
e^{i(1-\sigma) k.X}) \nonumber \\
&& +\int^1_0 d\sigma_1 
\int^1_{\sigma_1} d\sigma_2 \ i\del_{\sigma_1} w_\mu(\sigma_1) \ 
i\del_{\sigma_2} w_\nu(\sigma_2) \ {\rm Tr}(e^{i\sigma_1 k.X} \ X_\mu \ 
e^{i(\sigma_2-\sigma_1) k.X} X_\nu \ e^{i(1-\sigma_2) k.X}) \nonumber \\
&& +\cdots
\label{twofifteen}
\eea
The first term in the above expansion is just the operator that
couples to the tachyon. The second term vanishes because the operator
involved is independent of $\sigma$ due to the cyclic property of the
trace. Substituiting the mode expansion for $w(\sigma)$ in the third
term gives the following: 
\bea 
\sum^\infty_{n=1} \alpha^*_{n\mu} \alpha_{n\nu} \ (2\pi n)^2 \
\int^1_0 d\sigma \ e^{-2\pi in\sigma} \
{\rm Tr}(X_\mu \ e^{i\sigma k.X} \ X_\nu \ e^{i(1-\sigma) k.X})
\label{twosixteen}
\eea
From this expansion we may now read off the matrix model operators
that couple to specific modes of the closed string. As before, the
analogue of level matching condition emerges because of the cyclic
symmetry of the trace. As an example, the operators that couple to the
massless modes are given by the $n=1$ term in (\ref{twosixteen}). The
part of this term symmetric in $\mu, \nu$ has been shown in \cite{OO}
to give the stress tensor of the matrix model.

The general structure of the operators involved in the expansion
(\ref{twofifteen}) is easy to write down. The operator in the (r+1)th
term is
\bea
&& \int^1_0 d\sigma_1 \ e^{\mp 2\pi i n_1 \sigma_1 } 
\int^1_{\sigma_1} d\sigma_2 
\ e^{\mp 2\pi i n_2 \sigma_2} \cdots \int^1_{\sigma_{r-1}} d\sigma_r \ 
e^{\mp 2\pi i n_r \sigma_r} \nonumber \\
&& \hspace{10 ex} \times {\rm Tr}(e^{i\sigma_1 k.X} \ X_{\mu_1} \ 
e^{i(\sigma_2-\sigma_1)k.X} \ X_{\mu_2} \cdots e^{i(1-\sigma_r) k.X}) 
\label{twoseventeen}
\eea
The $\mp$ signs in the exponents correspond to mode $\alpha$ or its
conjugate $\alpha^*$. The above expression is non-vanishing only if
the signs are chosen such that the sum $\sum_l n_l$ for the positive
signs equals that for the negative signs \footnote{In the operators
  that appear in the expansion this expression has to be appropriately
  symmetrized by permuting the indices and the mode exponentials. The
  statement here holds only for the symmetrized operators like, for
  example, in (\ref{fivesix}).}. This is the analogue of the level
matching condition for the general term in the expansion
(\ref{twofifteen}).

It is easy to see that the operators that appear in the expansion of
the Wilson line $\tilde W_C$ are invariant under reparametrizations,
even though the modes of the periodic function that characterizes the
contour $C$ transform in a complicated way.  It is then natural to ask as
to how the sum in (\ref{twofifteen}) still manages to be
reparametrization invariant, as it must, since $\tilde W_C$ is by
definition reparametrization invariant. To see how this comes about,
let us consider a non-trivial example in detail.

The change in the harmonic function $w(\sigma)$ under
reparametrizations is given by the expression in (\ref{threethree}).
Consider now the first non-trivial term in
the expansion (\ref{twofifteen}), namely, the third term in this
equation.  It is easy to see that the change in this term due to the
first term in the variation of $w$ vanishes by itself. Its
variation due to the second term in the variation of $w$ is
given by \footnote{Here, a dot represents a derivative with respect
to the argument.}
\bea
&& \int^1_0 d\sigma_1 \int^1_{\sigma_1} d\sigma_2 \ 
\{\del_{\sigma_1}(-i\epsilon(\sigma_1) \ \dot w_\mu (\sigma_1)) \ 
i\dot w_\nu(\sigma_2)+i\dot w_\mu(\sigma_1) \ 
\del_{\sigma_2}(-i\epsilon(\sigma_2) \ \dot w_\nu(\sigma_2))\} 
\nonumber \\  
&& \hspace{20 ex} \times {\rm Tr}(e^{i\sigma_1 k.X} \ X_\mu \ 
e^{i(\sigma_2-\sigma_1) k.X} X_\nu \ e^{i(1-\sigma_2) k.X}) 
\label{twoeighteen}
\eea
This is clearly non-vanishing and hence for consistency must be
cancelled in the variation of the sum (\ref{twofifteen}). It turns out
that the variation of the next term in this sum under the first term
in the variation of $w$ does not vanish. In fact, the result is
a contribution that precisely cancels the above contribution, as we
shall see now.

The next term in the sum (\ref{twofifteen}) is
\bea
&& \int^1_0 d\sigma_1 \int^1_{\sigma_1} d\sigma_2 \int^1_{\sigma_2} d\sigma_3 \
i\dot w_\mu(\sigma_1) \ i\dot w_\nu(\sigma_2) \
i\dot w_\lambda(\sigma_3) \nonumber \\
&& \hspace{10 ex} \times {\rm Tr}(e^{i\sigma_1 k.X} \ X_\mu \ 
e^{i(\sigma_2-\sigma_1) k.X} X_\nu \ e^{i(\sigma_3-\sigma_2) k.X} X_\lambda \
 e^{i(1-\sigma_3) k.X})
\label{twonineteen}
\eea
The change in this under the first term in the variation of $w$ is
\bea
&& \int^1_0 d\sigma_1 \int^1_{\sigma_1} d\sigma_2 \int^1_{\sigma_2} d\sigma_3 \
\{(-i\dot \epsilon(\sigma_1)k_\mu) \ i\dot w_\nu(\sigma_2) \
i\dot w_\lambda(\sigma_3)+i\dot w_\mu(\sigma_1) \
(-i\dot \epsilon(\sigma_2)k_\nu) \ i\dot w_\lambda(\sigma_3) 
\nonumber \\ 
&& +i\dot w_\mu(\sigma_1) \ 
i\dot w_\nu(\sigma_2) \ (-i\dot \epsilon(\sigma_3)k_\lambda)\} \ 
{\rm Tr}(e^{i\sigma_1 k.X} \ X_\mu \ 
e^{i(\sigma_2-\sigma_1) k.X} X_\nu \ e^{i(\sigma_3-\sigma_2) k.X} X_\lambda \
e^{i(1-\sigma_3) k.X}) \nonumber \\
\label{twotwenty}
\eea
Now, partially integrating the derivative on $\epsilon$ and using
$\epsilon(0)=\epsilon(1)=0$, we find that the middle term in the curly
brackets above gives two contributions, which combine with the
contributions of the first and the last terms to give
\bea
&& \int^1_0 d\sigma_1 \int^1_{\sigma_1} d\sigma_2 \
\{(-i\epsilon(\sigma_1) \ \dot w_\mu (\sigma_1)) \ 
i\dot w_\nu(\sigma_2) \
\del_{\sigma_1}{\rm Tr}(e^{i\sigma_1 k.X} \ X_\mu \ 
e^{i(\sigma_2-\sigma_1) k.X} X_\nu \ e^{i(1-\sigma_2) k.X}) \nonumber \\
&& \hspace{10 ex} +i\dot w_\mu(\sigma_1) \
(-i\epsilon(\sigma_2) \ \dot w_\nu(\sigma_2)) \
\del_{\sigma_2}{\rm Tr}(e^{i\sigma_1 k.X} \ X_\mu \ 
e^{i(\sigma_2-\sigma_1) k.X} X_\nu \ e^{i(1-\sigma_2) k.X})\} \nonumber \\
\label{twotwentyone}
\eea
Partially integrating the derivatives and again using that
$\epsilon(0)=\epsilon(1)=0$, we find that this precisely cancels the
contribution in (\ref{twoeighteen}).

The above calculation illustrates the general mechanism by which the
sum in (\ref{twofifteen}) manages to be reparametrization invariant.
The structure of the operators that appear in the sum is such that the
contribution of the term at a given order under the second term in the
variation of $w$ is cancelled by the contribution of the next
higher order term due to the first term in the variation of $w$. 
In this respect the third term in (\ref{twofifteen}) is rather
special. For this term there is no lower order term whose contribution
would cancel its contribution under the $-\epsilon k$ term in the
variation of $w$. This contribution must, therefore, vanish by
itself. This ensures the existence of an infinite number of operators, 
given by the expression
\bea
O^{(n)}_{\mu\nu}(k)=\int^1_0 d\sigma \ 
e^{-2\pi in\sigma} \ {\rm Tr}(X_\mu \ e^{i\sigma k.X} \ X_\nu \ 
e^{i(1-\sigma) k.X}), \ n=1,2, \cdots, 
\label{twotwentytwo}
\eea
which satisfy the conservation equation
\bea
k_\mu O^{(n)}_{\mu\nu}(k)=0.
\label{twotwentythree}
\eea
The proof of this conservation equation is trivial and completely
kinematical.

The part of the operator for $n=1$ which is symmetric in the indices
$\mu, \nu$ is just the stress tensor of the matrix model. As noted in
\cite{OO}, the proof of the conservation of the stress tensor does not
require the equation of motion of the matrix model to be satisfied.
This is because in the zero slope limit conformal invariance of disc
amplitudes does not impose any restrictions on the gauge field. Here
we have seen this conservation equation emerging as a direct
consequence of the reparametrization invariance of the Wilson line.
From the space-time point of view, the conservation of the stress
tensor is equivalent to the decoupling of the longitudinal mode of the
graviton. The conservation equation (\ref{twotwentythree}) for
operators with $n \geq 2$ might at first seem surprising. However, it
turns out that this is not only required by reparametrization
invariance, as discussed above, but also because of some unexpected
identities satisfied by higher tensor operators. This will be
discussed in more detail at the end of the next section.
  
\section{Disc amplitudes and Wilson line operators}

In this section we would like to provide further evidence in support
of our suggestion that the matrix model operators (\ref{twoseventeen})
couple to different modes of the closed string. This evidence comes
from explicit calculations of disc amplitudes with the insertion of a
single closed string vertex operator, in the appropriate zero slope
limit. Since we are interested in the (bosonic) matrix model for
D-instantons, we have no directions parallel to the brane, and so the
calculations involve only the scalar fields and no gauge fields.

We wish to calculate disk amplitudes of the type
\bea
<V(z) \ {\rm Tr}({\rm P~exp}\{i\int^{+\infty}_{-\infty}dt \ \Phi_\mu 
i\del_\perp x^\mu(t)\})>
\label{fiveone}
\eea
Here $V(z)$ is a closed string vertex operator, the trace is over the
matrix $\Phi_\mu$, representing directions perpendicular to the
multiple D-instantons that we are considering, and $\del_\perp$ is a
derivative in the direction transverse to the boundary of the
world-sheet. With Dirichlet boundary conditions on $x^\mu$, the
propagator is given by
\bea
<x^\mu(z) \ x^\nu(w)>=-\alpha'g^{\mu\nu}\{{\rm log}|z-w|-{\rm log}|z-\bar w|\}.
\label{fivetwo}
\eea
The calculations now proceed similarly to those done in \cite{OO} for
the bosonic string.

Consider first the closed string tachyon vertex operator
$V(z)=e^{ik.x(z)}$. Using (\ref{fivetwo}) we get
\bea
&& <e^{ik.x(z)} \ {\rm Tr}({\rm P~exp}(i\int^{+\infty}_{-\infty}dt \ 
\Phi_\mu i\del_\perp x^\mu(t)))> \nn \\
&& \sim {\rm Tr}(e^{2\pi i \alpha' k.\Phi}) \nn \\
&& \equiv {\rm Tr}(e^{ik.X}).
\label{fivethree}
\eea
Here we have used that
\bea
i\del_\perp \{{\rm log}|z-t|-{\rm log}|\bar z-t|\} &=& {(z-\bar z) 
\over (z-t)(\bar z-t)} \nn \\
& \equiv & 2 \pi i \ \del_t \tau(t, z),
\label{fivefour}
\eea
 where, for a fixed value of $z$, the function ${1 \over 2 \pi i}{\rm
  log}({t-z \over t-\bar z}) \equiv \tau(t, z)$ is monotonic in $t$
and satisfies $\tau(\infty, z)-\tau(-\infty, z)=1$.  In the last step
in (\ref{fivethree}) we have used the definition $X \equiv 2\pi
\alpha' \Phi$ to match with the expressions in the previous section.
Also, we have ignored other factors since
we are only interested in that part of the result of the computation
which involves the matrix model variables.

We next consider the vertex operators at the massless level of the closed 
string, $V^{\mu\nu}(z)=\del x^\mu \bar\del x^\nu e^{ik.x(z)}$. Using
\bea
<\del x^\mu(z) \ i\del_\perp x^\nu(t)>=-{\alpha' g^{\mu\nu} \over
(z-t)^2} &=& -{2 \pi i \alpha' g^{\mu\nu} \over (z-\bar z)} \
e^{-2\pi i \tau(t, z)} \ \del_t \tau(t, z), \nn \\
<\bar \del x^\mu(\bar z) \ i\del_\perp x^\nu(t)>={\alpha' g^{\mu\nu} \over
(\bar z-t)^2} &=& {2 \pi i \alpha' g^{\mu\nu} \over (z-\bar z)} \
e^{2\pi i \tau(t, z)} \ \del_t \tau(t, z),
\label{fivefive}
\eea
we get
\bea
&& <\del x^\mu \bar\del x^\nu e^{ik.x(z)} \ {\rm Tr}({\rm P~exp}
\{i\int^{+\infty}_{-\infty}dt \ \Phi_\mu i\del_\perp x^\mu(t)\})> \nn \\
& \sim &
\int^1_0 d\tau_1 \ e^{-2\pi i \tau_1} \ \int^1_{\tau_1} d\tau_2  \ 
e^{2\pi i \tau_2} \ {\rm Tr}(e^{i \tau_1 k.X} \ X^\mu \
e^{i (\tau_2-\tau_1) k.X} \ X^\nu \  e^{i(1-\tau_2)k.X}) \nn \\
&& + \int^1_0 d\tau_1 \ e^{2\pi i \tau_1} \ \int^1_{\tau_1} d\tau_2  \ 
e^{-2\pi i \tau_2} \ {\rm Tr}(e^{i \tau_1 k.X} \ X^\nu \
e^{i (\tau_2-\tau_1) k.X} \ X^\mu \  e^{i(1-\tau_2)k.X}).
\label{fivesix}
\eea

At higher levels, there are several different types of closed string
vertex operators. For example, at the next level there are the
following three different types of vertex operators,
\bea
V_1^{\mu\nu\lambda\rho}(z) &=& \del x^\mu \del x^\nu \bar\del x^\lambda 
\bar\del x^\rho e^{ik.x(z)}, \nn \\
V_2^{\mu\nu\lambda}(z) &=& \del x^\mu \del x^\nu \bar\del^2 x^\lambda 
e^{ik.x(z)}, \nn \\
V_3^{\mu\nu}(z) &=& \del^2 x^\mu \bar\del^2 x^\nu e^{ik.x(z)}.
\label{fiveseven}
\eea
Inserting the first of these in (\ref{fiveone})and doing the 
computations as above, we get
\bea
&& <V_1^{\mu\nu\lambda\rho}(z) \ {\rm Tr}({\rm P~exp}
\{i\int^{+\infty}_{-\infty}dt \ \Phi_\mu i\del_\perp x^\mu(t)\})> \nn \\
& \sim &  
\int^1_0 d\tau_1 \ e^{-2\pi i \tau_1} \ \int^1_{\tau_1} d\tau_2  \ 
e^{-2\pi i \tau_2} \ \int^1_{\tau_2} d\tau_3  \ e^{2\pi i \tau_3} \ 
\int^1_{\tau_3} d\tau_4  \ e^{2\pi i \tau_4} \nn \\
&& \times {\rm Tr}(e^{i \tau_1 k.X} \ 
X^\mu \ e^{i (\tau_2-\tau_1) k.X} \ X^\nu \ 
e^{i (\tau_3-\tau_2) k.X} \ X^\lambda \ e^{i (\tau_4-\tau_3) k.X} \
X^\rho \ e^{i(1-\tau_4)k.X}) \nn \\
&& + \cdots
\label{fiveeight}
\eea
The dots stand for other terms which are obtained, like the
second term on the right hand side of (\ref{fivesix}), by permuting
the indices on the $X$'s together with the signs in the mode
exponentials. Similarly, for the other two operators in
(\ref{fiveseven}), using (\ref{fivefive}) and the identity
$1/(t-z)=(e^{-2\pi i \tau}-1)/(z-\bar z)$, we get
\bea
&& <V_2^{\mu\nu\lambda}(z) \ {\rm Tr}({\rm P~exp}
\{i\int^{+\infty}_{-\infty}dt \ \Phi_\mu i\del_\perp x^\mu(t)\})> \nn \\
& \sim &  
\int^1_0 d\tau_1 \ e^{-2\pi i \tau_1} \ \int^1_{\tau_1} d\tau_2  \ 
e^{-2\pi i \tau_2} \ \int^1_{\tau_2} d\tau_3  \ e^{4\pi i \tau_3} \nn \\
&& \times {\rm Tr}(e^{i \tau_1 k.X} \ 
X^\mu \ e^{i (\tau_2-\tau_1) k.X} \ X^\nu \ 
e^{i (\tau_3-\tau_2) k.X} \ X^\lambda \ e^{i(1-\tau_3)k.X}) \nn \\
&& + \cdots, 
\label{fivenine}
\eea
and 
\bea
&& <V_3^{\mu\nu}(z) \ {\rm Tr}({\rm P~exp}
\{i\int^{+\infty}_{-\infty}dt \ \Phi_\mu i\del_\perp x^\mu(t)\})> \nn \\
& \sim &  
\int^1_0 d\tau_1 \ e^{-4\pi i \tau_1} \ \int^1_{\tau_1} d\tau_2  \ 
e^{4\pi i \tau_2} \ {\rm Tr}(e^{i \tau_1 k.X} \ 
X^\mu \ e^{i (\tau_2-\tau_1) k.X} \ X^\nu \ e^{i(1-\tau_2)k.X}) \nn \\
&& +
\int^1_0 d\tau_1 \ e^{-2\pi i \tau_1} \ \int^1_{\tau_1} d\tau_2  \ 
e^{2\pi i \tau_2} \ {\rm Tr}(e^{i \tau_1 k.X} \ 
X^\mu \ e^{i (\tau_2-\tau_1) k.X} \ X^\nu \ e^{i(1-\tau_2)k.X}) \nn \\
&& + \cdots   
\label{fiveten}
\eea

We see that in all the cases above, the matrix model part of the
result has the generic form given in (\ref{twoseventeen}). It should
also be noted that for the vertex operator $V_3$, the result of the
calculation gives a linear sum of matrix model expressions of the type
in (\ref{twoseventeen}). The above calculations can be easily
generalized to higher levels and it turns out that they have the same
generic features.

Before we end this section, let us come back to the conserved
operators $O^{(n)}_{\mu\nu}(k)$ identified in the previous section.
From the above calculations, we see that the operator for $n=2$ is
dual to the closed string mode given by the vertex operator $V_3$,
and so on. Conservation of these may, therefore, seem puzzling since
$k_\mu V_3^{\mu\nu}$ is not a total world-sheet derivative. However,
the difference from a total derivative is essentially given by $k_\mu
k_\nu V_2^{\mu\nu\lambda}$. The matrix model operator dual to this
operator vanishes, as can easily be seen from the above calculations.
It is these type of unexpected identities satisfied by the matrix
model operators that are responsible for the infinite set of conserved
operators that we have identified here.

\section{Symmetries of the generating functional of Wilson line correlators}

In the previous sections we have established that the correlation
functions of the gauge theory operators that appear in the harmonic
expansion of a generic Wilson line $W_C$ are invariant under
reparametrizations, even though the modes $\alpha_n$ of the periodic 
function $w(\sigma)$ change in a complicated way. In this
section we will study a non-trivial consequence of this fact. 

Let us add to the non-commutative gauge theory action source terms for
all the operators that appear in the harmonic expansion:
\bea
\int d^Dk
\sum_{n_1,n_2, \cdots=1}^\infty \ \sum_{ m_1,m_2, \cdots=1}^\infty
\Phi^{n_1,n_2,... \ ; \ m_1,m_2,...}_{\mu_1,\mu_2,... \ ; \ 
\nu_1,\nu_2,...}(k) \
O^{n_1,n_2,... \ ; \ m_1,m_2,...}_{\mu_1,\mu_2,... \ ; \ 
\nu_1,\nu_2,...}(k)
\label{fourone}
\eea 
A more elegant way of writing this is the following. Let us introduce
the ``closed string field''  $\Phi_k[w] \equiv \Phi_k(\alpha_1,
\alpha_2, \cdots)$ \footnote{The condition $\Phi^*_{-k}=\Phi_k$ is 
needed to ensure that the source terms in the action are real}
and define the sources as its ``moments'':
\bea
&& \Phi^{n_1,n_2,... \ ; \ m_1,m_2,...}_{\mu_1,\mu_2,... \ ; \ 
\nu_1,\nu_2,...}(k) \nonumber \\
&& =\int \prod_{n=1}^\infty d\alpha_n \ d\alpha_n^* \
\alpha_{n_1\mu_1} \alpha_{n_2\mu_2} \cdots \
\alpha_{m_1\nu_1}^* \alpha_{m_2\nu_2}^* \cdots \
\Phi_k(\alpha_1, \alpha_2, \cdots).
\label{fourtwo}
\eea
We may then rewrite (\ref{fourone}) in the compact form 
\bea
\int d^Dk \int \prod_{n=1}^\infty d\alpha_n \ d\alpha_n^* \
\Phi_k(\alpha_1, \alpha_2, \cdots) \ W_{C_0}(\alpha_1, \alpha_2, \cdots)
\label{fourthree}
\eea
where we have explicitly indicated the parametrization of the contour
$C$ in terms of the straight line $C_0$ (which is given by the
starting point $y_0(0)$ and the the momentum $k$) and the modes of the
periodic function $w$.

In the above we have assumed that the Wilson line is expanded in terms
of the modes of the periodic function $w$. We could equally well have
used the parametrization of the contour $C$ in terms of $C_0'$ and
$w'$ and then the Wilson line would be expanded in terms of the modes
of $w'$. Replacing $W_{C_0}(\alpha_1, \alpha_2, \cdots)$ by
$W_{C_0'}(\alpha_1', \alpha_2', \cdots)$ in (\ref{fourthree}) has the
effect of changing the sources because the primed modes are
non-trivial functions of the unprimed ones and because the function
$\Phi_k(\alpha_1, \alpha_2, \cdots)$ has not been changed. These
transformed sources, $\Phi'$, are now coupled to the primed operators
$O'$ discussed in the previous section. However, since the correlation
functions of $O'$ are identical to those of the operators $O$, it
follows that the generating functional of the correlators of these
operators, which is given by
\bea 
Z[\Phi]=<{\rm exp}\{\int d^Dk \int \prod_{n=1}^\infty d\alpha_n \
d\alpha_n^* \ \Phi_k(\alpha_1, \alpha_2, \cdots) \ 
W_{C_0}(\alpha_1, \alpha_2, \cdots)\}>,
\label{fourfour}
\eea
remains unchanged, i.e.
\bea
Z[\Phi']=Z[\Phi].
\label{fourfive}
\eea

How are the transformed sources $\Phi'$ related to $\Phi$? To see this
it is useful to look at some explicit examples. The first non-trivial
example is provided by the sources which couple to gauge theory
operators that are dual to the massless closed string modes. In terms
of the definition in (\ref{fourtwo}) these sources are given by
\bea
\Phi^{1;1}_{\mu;\nu}(k)
=\int \prod_{n=1}^\infty d\alpha_n \ d\alpha_n^* \
\alpha_{1\mu}\alpha_{1\nu}^* \ \Phi_k(\alpha_1, \alpha_2, \cdots)
\label{foursix}
\eea
After a reparametrization of the contour $C$ these sources transform to
\bea
\Phi_{\mu;\nu}^{1;1}(k)'
=\int \prod_{n=1}^\infty d\alpha_n \ d\alpha_n^* \
\alpha_{1\mu}'\alpha_{1\nu}'^* \ \Phi_k(\alpha_1, \alpha_2, \cdots)
\label{foursixa}
\eea
Using (\ref{threeseven}) in this and retaining terms only upto first 
order in $\epsilon$, we get
\bea
\Phi_{\mu;\nu}^{1;1}(k)'-\Phi^{1;1}_{\mu;\nu}(k)
&=& -k_\mu \epsilon_1 \Phi^{;1}_{;\nu}(k)
-k_\nu \epsilon_1^* \Phi^{;1}_{;\mu}(-k)
+\{4\pi i \epsilon_1^* 
\Phi^{2;1}_{\mu;\nu}(k) \nn \\
&& -2\pi i \epsilon_2 \Phi^{;1,1}_{;\mu,\nu}(k) 
+6\pi i \epsilon_2^* \Phi^{3;1}_{\mu;\nu}(k)+{\rm c.c.}\}+\cdots,
\label{fourseven}
\eea
where,
\bea
\Phi^{;1}_{;\mu}(k)=\int \prod_{n=1}^\infty d\alpha_n \
d\alpha_n^* \ \alpha_{1\mu}^* \ \Phi_k(\alpha_1, \alpha_2, \cdots),
\label{foureight}
\eea
follows from the definition (\ref{fourtwo}) and similarly for
$\Phi^{;1,1}_{;\mu,\nu}(k)$ etc. The `c.c.' in (\ref{fourseven}) stands
for complex conjugation together with $\mu \leftrightarrow \nu$ and $k
\rightarrow -k$ and the dots represent terms involving $\epsilon_3$
and higher modes of $\epsilon(\sigma)$.

Let us first focus on terms proportional to $\epsilon_1$ on the right
hand side of (\ref{fourseven}). We see that the symmetric part of the
source $\Phi_{\mu;\nu}^{1;1}(k)$, which couples to the gauage theory
operator dual to the closed string graviton, transforms just like the
linearised transformation of metric under general coordinate
transformations. Similarly, the antisymmetric part of
$\Phi_{\mu;\nu}^{1;1}(k)$ transforms just like the antisymmetric
tensor field under gauge transformations. The presence of these
space-time symmetries implies that the longitudinal modes of the
operators dual to graviton and antisymmetric tensor field decouple. In
string theory, this corresponds to the decoupling of the spurious states
\bea
L_{-1} \tilde a_{1\mu}^\dagger|0;k>, \ \ \tilde L_{-1} a_{1\mu}^\dagger|0;k>,
\label{fournine}
\eea 
where $a_n, \tilde a_n$ are the left and right moving oscillator modes
of the closed string and the $L_n, \tilde L_n$ are the corresponding
virasoro generators. From the transformation of 
$\Phi^{1;1}_{\mu;\nu}(k)$ we see that the non-commutative gauge 
theory/matrix model analogues of these states are the fields 
$\Phi^{;1}_{;\mu}(k)$ and $\Phi^{1;}_{\mu;}(k)$. 

At level two we have several sources,
$\Phi^{1,1;1,1}_{\mu,\nu;\lambda\rho}(k)$,
$\Phi^{1,1;2}_{\mu,\nu;\lambda}(k)$, $\Phi^{2;2}_{\mu;\nu}(k)$ and
their congugates. Using
\bea
\alpha_2'=\alpha_2-k\epsilon_2+4\pi i \epsilon_0 \alpha_2+
2\pi i \epsilon_1 \alpha_1+2\pi i \sum^\infty_{n=1}
((n+2)\epsilon_n^* \alpha_{n+2}-n\epsilon_{n+2} \alpha_n^*),
\label{fourten}
\eea
we find that the source $\Phi^{2;2}_{\mu;\nu}(k)$ transforms as 
\bea
\Phi^{2;2}_{\mu;\nu}(k)'-\Phi^{2;2}_{\mu;\nu}(k) &=& -k_\mu \epsilon_2
\Phi^{;2}_{;\nu}(k)-k_\nu \epsilon_2^* \Phi^{;2}_{;\mu}(-k)
+\{2\pi i \epsilon_1 \Phi^{1;2}_{\mu;\nu}(k) \nn \\
&& +6\pi i \epsilon_1^* \Phi^{3;2}_{\mu;\nu}(k)
+8\pi i \epsilon_2^* \Phi^{4;2}_{\mu;\nu}(k)+{\rm c.c.}\}+\cdots
\label{foureleven}
\eea
Similarly,
\bea
\Phi^{1,1;2}_{\mu,\nu;\lambda}(k)'-\Phi^{1,1;2}_{\mu,\nu;\lambda}(k) &=&
(-k_\mu \epsilon_1 \Phi^{1;2}_{\nu;\lambda}(k)+4\pi i \epsilon_1^*
\Phi^{2,1;2}_{\mu,\nu;\lambda}(k)-2\pi i \epsilon_2
\Phi^{1;1,2}_{\nu;\mu,\lambda}(k)+6\pi i \epsilon_2^*
\Phi^{3,1;2}_{\mu,\nu;\lambda}(k)) \nn \\
&& +(\mu \leftrightarrow \nu)
-k_\lambda \epsilon_2^* \Phi^{1,1;}_{\mu,\nu;}(k)-6\pi i \epsilon_1
\Phi^{1,1;3}_{\mu,\nu;\lambda}(k)-8\pi i \epsilon_2
\Phi^{1,1;4}_{\mu,\nu;\lambda}(k) \nn \\
&& -2\pi i \epsilon_1^* \Phi^{1,1;1}_{\mu,\nu;\lambda}(k)+\cdots
\label{fourtwelve}
\eea
One can also similarly work out the transformation of the source
$\Phi^{1,1;1,1}_{\mu,\nu;\lambda\rho}(k)$. It is easy to see that in
all the cases, the gauge transformation due to $\epsilon_1$ at this
level is consistent with the decoupling of operators whose string theory
analogues are the spurious states 
\bea
L_{-1} a_{1\mu}^\dagger \tilde a_{2\nu}^\dagger |0;k>, \ \ 
L_{-1} a_{1\mu}^\dagger \tilde a_{1\nu}^\dagger 
\tilde a_{1\lambda}^\dagger |0;k>, \ \
\tilde L_{-1} a_{2\mu}^\dagger \tilde a_{1\nu}^\dagger |0;k>, \ \
\tilde L_{-1} a_{1\mu}^\dagger a_{1\nu}^\dagger \tilde 
a_{1\lambda}^\dagger |0;k>,
\label{fourthirteen}
\eea
etc. We should note here that the first term in the curly brackets on
the right hand side of (\ref{fourseven}) is also of this type. This is
because, as we observed in the computation of the disc amplitude
involving the vertex operator $V_3$ in the last section, a closed
string vertex operator at a given level is in general dual to a linear
combination of gauge theory operators upto that level \footnote{Here,
  by the level of a gauge theory operator we mean the sum of positive
  (or negative) numbers in the mode exponentials. For example, the
  level of the operator $O^{(n)}_{\mu\nu}$ given in
  (\ref{twotwentytwo}) is $n$.}.

We can also now interpret the gauge transformation due to $\epsilon_2$
in terms of string theory. From the above transformations, it is clear
that this corresponds to the spurious states
\bea
L_{-2} \tilde a_{2\mu}^\dagger |0;k>, \ \ 
\tilde L_{-2} a_{2\mu}^\dagger |0;k>,
\label{fourfourteen}
\eea
etc.

At higher levels and for higher modes of $\epsilon(\sigma)$ a similar
pattern is confirmed and one finds that the gauge transformations of
the sources are in one-to-one correspondence with the spurious states
of string theory. In string theory this gauge symmetry allows one to
remove the entire tower of oscillator modes corresponding to a single
coordinate degree of freedom. The analogue of this here is the fact
that in the matrix model one can always diagonalize one of the
matrices by a unitary transformation, so the actual dynamical degrees
of freedom contain one less matrix. This fact is not so obvious in
terms of the operators that appear in the harmonic expansion of the
Wilson line, since these operators are manifestly invariant under
unitary transformations of the matrices. However, in the space-time
interpretation of these operators, in which each matrix is interpreted
as a coordinate, this fact reappears as a gauge symmetry. In this
sense it is natural and consistent to find that the gauge symmetry of
the generating functional $Z[\Phi]$ is identical to the gauge symmetry
of closed string theory.

\section{Discussion}

In this paper we have investigated the proposal that the harmonic
expansion of an open Wilson line, defined on a generic contour, around
a certain straight line contour, contains gauge theory operators dual
to {\it all} the modes of the closed string. This was motivated by the
observation that the first two operators in the expansion are dual to
the tachyon and the massless modes of the closed string. We argued
that the gauge theory operators that appear in the harmonic expansion
satisfy an analogue of the level matching condition. Further evidence
for our proposal came from some explicit computations of disc
amplitudes. We also discussed space-time gauge symmetries of the
generating functional of correlation functions of these operators. The
gauge symmetries are identical to what one finds in the field theory
of closed strings. We showed that the symmetries are a direct
consequence of the reparametrization invariance of the original Wilson
line.

An aspect of the present investigations that needs to be clarified is
the absence of any condition on the dimensions of the target space.
Presumably this is related to the fact that in the bosonic string
there is no condition on the background fields in the sigma-model at
the classical level. Our analysis here has also been classical. It is
conceivable that the formal reparametrization invariance of the Wilson
line is not realized in the generating functional defined in
(\ref{fourfour}) at the quantum level unless the sources satisfy
certain conditions, including a condition on the number of space-time
dimensions. In the supersymmetric case, ensuring space-time
supersymmetry, for example by ensuring kappa-symmetry in the
Green-Schwarz formalism, leads to a restriction on the background
fields already at the classical level. It would, therefore, be very
intersting to generalize the present analysis to the case of the
superstring.

Our analysis here has been essentially kinematical, relying only on
the symmetries of the system. Investigating its dynamics is clearly
necessary for further progress. For example, what role do
Schwinger-Dyson or Loop equations play in the story? These equations
represents dynamical constraints on the correlation functions of the
gauge theory operators. It would be interesting to find out what kind
of constraints they imply on the functional form of the generating
functional $Z[\Phi]$.

Another issue related to the quantum dynamics of the present system is
the following. In the representation theory of Virasoro algebra, the
critical dimension is associated with further reducibility due to the
emergence of null states. After eliminating the null states, we are
left with the light-cone gauge like physical degrees of freedom. We
need to identify the central charge associated to the quantum Virasoro
algebra for the Wilson lines to determine the critical dimension of
the closed string theory dual to it. Quantum dynamics of Wilson lines
certainly plays an important role here and in principle it can be
extracted from the Loop equations.

Finally, we point out that we may also regard (\ref{fourfour}) as the
partition function of bosonic version of the matrix model in weak
background fields \footnote{Various aspects of matrix models in
  background fields have been studied in
  \cite{D,DKO,DO,KT,TR1,KK,TR2,TR3,M,DNP}}. Actually since all possible
terms in the matrix theory action can be obtained from special sources
\footnote{For example, the commutator squared term in the matrix model
  action corresponds to the source $\Phi^{1;1}_{\mu;\nu}(k)
  \sim \del_{k_\mu} \del_{k_\nu} \delta^D(k)$. This follows
  from the identity Tr$[X_\mu,X_\nu]^2 \sim \del_{k_\mu} \del_{k_\nu} 
O^{1;1}_{\mu;\nu}(k)|_{k=0}$.}, the entire
action can be thought of as just matrix theory operators coupled to
sources. This is very similar to the world-sheet sigma model action
for the bosonic string in background fields. It would be interesting
to see how far this analogy holds. In particular, does quantum
consistency of the matrix model impose restriction on the sources? One
particular consistency condition that we should require of the quantum
theory is the reparametrization invariance or equivalently the gauge
invariance of (\ref{fourfour}). It would be very interesting to see
whether this condition is enough to recover the correct string
dynamics within this framework.

\end{document}